\documentclass[twocolumn,english,superscriptaddress,citeautoscript,showpacs,preprintnumbers,amsmath,amssymb,prb,floatfix,footinbib]{revtex4}
\usepackage[scaled=2]{helvet}
\usepackage[T1]{fontenc}
\usepackage[latin9]{inputenc}
\usepackage{color}
\definecolor{note_fontcolor}{rgb}{0.800781, 0.800781, 0.800781}
\usepackage{textcomp}
\usepackage{dsfont}
\usepackage{amsmath}
\usepackage{amssymb}
\usepackage{graphicx}
\usepackage{ulem}

\makeatletter

\@ifundefined{textcolor}{}
{%
 \definecolor{BLACK}{gray}{0}
 \definecolor{WHITE}{gray}{1}
 \definecolor{RED}{rgb}{1,0,0}
 \definecolor{GREEN}{rgb}{0,1,0}
 \definecolor{BLUE}{rgb}{0,0,1}
 \definecolor{CYAN}{cmyk}{1,0,0,0}
 \definecolor{MAGENTA}{cmyk}{0,1,0,0}
 \definecolor{YELLOW}{cmyk}{0,0,1,0}
}

\usepackage[scaled=2]{helvet}\usepackage{amstext}

\makeatletter
\@ifundefined{textcolor}{}{\definecolor{BLACK}{gray}{0}\definecolor{WHITE}{gray}{1}\definecolor{RED}{rgb}{1,0,0}\definecolor{GREEN}{rgb}{0,1,0}\definecolor{BLUE}{rgb}{0,0,1}\definecolor{CYAN}{cmyk}{1,0,0,0}\definecolor{MAGENTA}{cmyk}{0,1,0,0}\definecolor{YELLOW}{cmyk}{0,0,1,0}}

\usepackage{dcolumn}
\usepackage{bm}
\usepackage{hyperref}
\hypersetup{colorlinks=true,linkcolor=red,citecolor=blue}
\makeatother

\usepackage{babel}

\makeatother

\usepackage{babel}
\begin{document}
\title{Magnetic excitations in double perovskite iridates La$_{2}$$\mathit{M}$IrO$_{6}$
($\mathit{M}$ = Co, Ni, and Zn) mediated by 3$\mathit{d}$-5$\mathit{d}$
hybridization}
\author{Wentao Jin}
\email{wtjin@buaa.edu.cn}

\affiliation{Department of Physics, University of Toronto, Toronto, Ontario, M5S
1A7, Canada}
\affiliation{School of Physics, Beihang University, Beijing 100191, China}
\author{Sae Hwan Chun}
\affiliation{Department of Physics, University of Toronto, Toronto, Ontario, M5S
1A7, Canada}
\affiliation{Pohang Accelerator Laboratory, Pohang, Gyeongbuk 37673, Republic of
Korea}
\author{Jungho Kim}
\affiliation{Advanced Photon Source, Argonne National Laboratory, Argonne, Illinois
60439, USA}
\author{Diego Casa}
\affiliation{Advanced Photon Source, Argonne National Laboratory, Argonne, Illinois
60439, USA}
\author{Jacob P. C. Ruff}
\affiliation{Cornell High Energy Synchrotron Source, Cornell University, Ithaca,
New York 14853, USA}
\author{C. J. Won}
\affiliation{Department of Physics, Inha University, Incheon 22212, Republic of
Korea}
\affiliation{Laboratory for Pohang Emergent Materials and Max Plank POSTECH Center for Complex Phase Materials, Pohang University of Science and Technology, Pohang 37673, Republic of Korea}
\author{K. D. Lee}
\affiliation{Department of Physics, Inha University, Incheon 22212, Republic of
Korea}
\author{N. Hur}
\affiliation{Department of Physics, Inha University, Incheon 22212, Republic of
Korea}
\author{Young-June Kim}
\email{youngjune.kim@utoronto.ca}

\affiliation{Department of Physics, University of Toronto, Toronto, Ontario, M5S
1A7, Canada}

\date{\today}

\begin{abstract}
By performing resonant inelastic x-ray scattering (RIXS) measurements
at the Ir $\mathit{L_{\mathrm{3}}}$ edge, we have investigated the low-energy
elementary excitations in a series of double perovskite iridate single
crystals, La$_{2}$$\mathit{M}$IrO$_{6}$ ($\mathit{M}$ = Co, Ni,
and Zn). Almost dispersionless magnetic excitations at $\mathds{\sim}$ 42(6) meV and $\mathds{\sim}$ 35(5) meV have been observed in crystals containing magnetic 3$\mathit{d}$ ions, La$_{2}$CoIrO$_{6}$ and La$_{2}$NiIrO$_{6}$, respectively. 
In contrast, this low-energy magnetic
excitation is absent in La$_{2}$ZnIrO$_{6}$ in which the 3$\mathit{d}$
ions are non-magnetic, suggesting the importance of 3$\mathit{d}$-5$\mathit{d}$
hybridization in the magnetic properties of these double perovskite
iridates. The magnetic excitation is suppressed completely above the magnetic
ordering temperature, suggesting the inadequacy of using a simple spin Hamiltonian to describe magnetism of these materials. 
\end{abstract}
\maketitle

\section{Introduction}

Transition metal oxides with 4$\mathit{d}$ and 5$\mathit{d}$ elements
have attracted tremendous interest in the past decade, due to the
possible realization of exotic quantum states such as spin-orbital
assisted Mott insulators,\cite{KimBJ_08,Plumb_14} quantum spin liquids,\cite{Banerjee_16,Takagi_19}
and Weyl semimetals,\cite{Wan_11} arising from the delicate
interplay between the strong spin-orbit coupling (SOC), crystal-field
splitting, and Coulomb repulsion.\cite{Witczak-Krempa_14,Rau_16}
Especially, the largely extended 5$\mathit{d}$ orbitals render a
relatively weaker electronic correlation compared with that of the
3$\mathit{d}$ transition metal (TM), leading to various attempts
aiming to combine the strong correlation of the 3$\mathit{d}$ TM
and strong SOC of the 5$\mathit{d}$ TM by engineering 3$\mathit{d}$-5$\mathit{d}$
heterostructures.\cite{Matsuno_15,Nichols_16,Hao_18} Double perovskite
(DP) iridates with the formula of $\mathit{A_{\mathrm{2}}B}$IrO$_{6}$,
in which $\mathit{A}$ is a rare-earth or alkaline-earth metal and
$\mathit{B}$ is a 3$\mathit{d}$ TM, respectively, offer a unique
possibility to realize such a 3$\mathit{d}$-5$\mathit{d}$ combination
even within the bulk structure. The topic of $3d-5d$ coupled magnetism in DP structure has been the subject of extensive research efforts over the years. \cite{Cook_14,Morrow_13,Vasala_15,Plumb_13,Laguna-Marco_15,Vogl_18,LeeM_18,Yuan_18,Taylor_18,Granado_19,Bhowal_20,Yuan_21}

La$_{2}$$\mathit{M}$IrO$_{6}$ with either magnetic ($\mathit{M}$
= Mn, Co, Ni, Cu) or nonmagnetic ($\mathit{M}$ = Mg, Zn) $\mathit{M}^{2+}$
ions adopts the $\mathit{B}$-site ordered rock-salt structure with
a monoclinic symmetry (space group $\mathit{P}$2$_{1}$/$\mathit{n}$),\cite{Vasala_15,Currie_95}
which can be regarded as two interpenetrating simple
perovskite lattices La$\mathit{M}$O$_{3}$ and LaIrO$_{3}$ with
alternating $\mathit{M}^{2+}$O$_{6}$ and Ir$^{4+}$O$_{6}$ octahedra.
With the 5$\mathit{d}^{5}$ electronic configuration, the Ir $\mathit{t}$$_{\mathrm{2g}}$
manifold splits into a fully filled $\mathit{j_{\mathrm{eff}}}$
= 3/2 quartet and a half filled $\mathit{j_{\mathrm{eff}}}$
= 1/2 doublet in the presence of strong SOC. As a result, a spin-orbital
entangled $\mathit{j_{\mathrm{eff}}}$ = 1/2 Mott insulator
is expected as their electronic ground state.\cite{Cao_13} Such a
$\mathit{j_{\mathrm{eff}}}$ = 1/2 state associated with
the Ir$^{4+}$ ions on a quasi-face-centered cubic lattice can host
the long-sought Kitaev interactions,\cite{Kitaev_06,Jackeli_09,Kimchi_14}
which drives numerous theoretical and experimental studies in recent
years on the magnetism of DP iridates La$_{2}$$\mathit{M}$IrO$_{6}$.

In the case of La$_{2}$MgIrO$_{6}$ and La$_{2}$ZnIrO$_{6}$ with nonmagnetic $\mathit{M}^{2+}$ ions occupying the $\mathit{B}$-site,
the Ir$^{4+}$ sublattice was found to order in a collinear and canted antiferromagnetic
(AFM) structure below $\mathit{T_{\mathrm{N}}}$ = 12 K and 7.5 K, respectively,
with a $\mathit{k}$ = 0 propagation vector.\cite{Cao_13} The canted
antiferromagnetism in La$_{2}$ZnIrO$_{6}$ (LZIO) was modeled by a spin Hamiltonian incorporating both the Heisenberg exchange and the
Dzyaloshinskii\textendash Moriya interaction due to the octahedral titing and strong deviation
of the Zn-O-Ir angle from 180$^{\circ}$.\cite{Zhu_15} On the other hand, in the
presence of a magnetic $\mathit{M}^{2+}$ ion in La$_{2}$$\mathit{M}$IrO$_{6}$,
the magnetic ground state is more complicated due to the interplay
between 3$\mathit{d}$ and 5$\mathit{d}$ magnetism. For example,
La$_{2}$MnIrO$_{6}$ undergoes a ferromagnetic (FM) transition at
$\mathit{T_{\mathrm{C}}}$ = 130 K,\cite{Demazeau_94} whereas La$_{2}$CuIrO$_{6}$
exhibits an AFM transition into a noncollinear magnetic structure
with $\mathit{k}$ = (1/2, 0, 1/2).\cite{Manna_16} La$_{2}$CoIrO$_{6}$
(LCIO) display a ferrimagnetic order below $\mathit{T_{\mathrm{C}}}$ $\mathds{\sim}$
95 K as revealed by neutron powder diffraction.~\cite{Narayanan_10,LeeS_18} In this noncollinear magnetic order with $\mathit{k}$ = 0,
the moments of both the Co and Ir sublattice are ordered antiferromagnetically in the $\mathit{ac}$ plane and canted along the $\mathit{b}$
axis, giving rise to ferromagnetic moments.  In addition, La$_{2}$NiIrO$_{6}$
(LNIO) was recently determined to be a noncollinear antiferromagnet
with $\mathit{k}$ = (1/2, 1/2, 0), showing only the $\mathit{a}$-
and $\mathit{c}$- spin components with simultaneous ordering of Ni$^{2+}$
and Ir$^{4+}$ moments.\cite{Ferreira_21}

Although many experimental studies have been carried out to understand the static
magnetic order in the DP iridates La$_{2}$$\mathit{M}$IrO$_{6}$,
information about the spin dynamics have been scarce. For LZIO, a recent inelastic neutron scattering study found
gapped spin-wave excitations with very weak dispersion, suggesting
strong exchange anisotropies and a possible dominant Kitaev interaction
associated with the monoclinically distorted DP lattice.\cite{Aczel_16}
The recent combined nuclear magnetic resonance
(NMR) and electron spin resonance (ESR) measurements have revealed
pronounced quasi-static spin correlations in both LZIO and LCIO persisting
at temperatures far above their magnetic phase transitions, signifying
substantial magnetic frustration inherently existing in these DP iridates
composed of two intersecting fcc sublattices.\cite{Iakovleva_18}
However, it is still not clear how the magnetism is translated between
the 3$\mathit{d}$ magnetic TM and 5$\mathit{d}$ Ir$^{4+}$ ions,
and what roles the 3$\mathit{d}$ magnetic TM ions is playing in the
spin dynamics of the Ir sublattice. In this paper, the low-energy
magnetic excitations of Ir$^{4+}$ pseudospins in a series of DP iridates,
La$_{2}$$\mathit{M}$IrO$_{6}$ ($\mathit{M}$ = Co, Ni, and Zn),
are systematically studied using resonant inelastic x-ray scattering
(RIXS) measurements at the Ir $\mathit{L}_{3}$ edge. A large excitation
gap of $\mathds{\sim}$ 40 meV was observed in both LCIO and LNIO,
but not in LZIO, indicating the important role of the 3$\mathit{d}$
magnetic ions in mediating the spin exciations of the Ir sublattice.

\section{Experimental Details}

RIXS experiments were performed at the 27 ID-B beamline of the Advanced
Photon Source (APS) at Argonne National Laboratory using the MERIX
instrumentation.\cite{Gog_09} A double-bounce diamond(111) primary
monochromator, a channel-cut Si(844) secondary monochromator, and
a spherical (2 m radius) diced Si(844) analyzer crystal were used
to obtain an overall energy resolution of $\mathds{\sim}$ 25 meV
{[}full width at half-maximum (FWHM){]} at the incident photon energy
$\mathit{E_{i}}$ = 11.214 keV (near the Ir $\mathit{L_{\mathrm{3}}}$
edge, 2$\mathit{p}$ $\mathds{\rightarrow}$ 5$\mathit{d}$). The
incident beam was focused to a size of 40 \texttimes{} 20 $\mu$m$^{2}$
($\mathit{H}$ \texttimes{} $\mathit{V}$) at the sample position.
A MYTHEN strip detector was used, and the horizontal scattering geometry
with the scattering angle 2$\theta$ close to 90$^{\circ}$ was adopted to minimize
the elastic background intensity. The sample temperature was controlled
between 5 and 300 K in a closed-cycle refrigerator.

Single crystals of La$_{2}$$\mathit{M}$IrO$_{6}$ ($\mathit{M}$ = Co, Ni, and Zn) were synthesized using the PbO-PbF$_{2}$ flux method. The polycrystalline samples of La$_{2}$$\mathit{M}$IrO$_{6}$ were firstly prepared by solid-state reaction.  The stoichiometric mixtures of pre-dried La$_{2}$O$_{3}$, Co$_{3}$O$_{4}$/NiO/ZnO, and IrO$_{2}$ powders were ground, pelletized, and sintered at 1150 \textcelsius{} twice with an intermediate grinding. The obtained polycrystalline sample La$_{2}$$\mathit{M}$IrO$_{6}$ were then mixed with the PbO-PbF$_{2}$ flux inside a Pt crucible covered by an AlO$_{2}$OO$_{3}$ thermal mass. The mixture were heated up to 1300 \textcelsius{} for 6 h followed by a slow cooling procedure of 3-5 \textcelsius{}/h to 1000 \textcelsius{}, after which millimeter-size crystals were obtained and seperated from the bulk. 

Using the monoclinic notation, the crystals used in the RIXS experiments
were polished with the {[}0, 1, 0{]}, {[}0, 0, 1{]} and {[}1, 1, 0{]}
directions parallel to the surface normals for LCIO, LNIO and LZIO,
respectively. Structural characterizations of the LCIO 
and LNIO single crystals were performed at the A2
beamline of the Cornell High Energy Synchrotron Source (CHESS) using
high-energy x-ray diffraction with $\mathit{E_{i}}$ = 60 keV, at
both $\mathit{T}$ = 280 K and $\mathit{T}$ = 100 K. The dc magnetization
data of the same crystals used in synchrotron x-ray measurements were
collected using a Quantum Design magnetic property measurement system
(MPMS).

As shown in Figure S1 in the Supplemental Material,\cite{SM} at room temperature,
all the diffraction spots of the La$_{2}$$\mathit{M}$IrO$_{6}$
($\mathit{M}$ = Co and Ni) single crystals fit well with a monoclinic
lattice ($\mathit{a}$ $\mathds{\approx}$ 5.58 \AA{}, $\mathit{b}$ $\mathds{\approx}$
5.67 \AA{}, $\mathit{c}$ $\mathds{\approx}$ 7.91 \AA{}, $\beta$ $\mathds{\approx}$
89.9$^{\circ}$) with the formation of multiple twins (note that $\mathit{d_{\mathrm{110}}}$
$\mathds{\approx}$ $\mathit{d_{\mathrm{002}}}$), evidenced by the
appearances of (odd/2, odd/2, 0) reflections. This is consistent with
the previously reported monoclinic symmetry for DP iridates.\cite{Currie_95}
Upon cooling down to 100 K, the diffraction patterns do not show any
visible change for both compounds, indicating the robustness of the
monoclinic structure with decreasing temperature. Therefore, throughout
this paper we will use the monoclinic notation to describe their crystal
structure. The sharpness of all diffraction spots confirms the high
quality of the single-crystal samples. Although we did not carry out single-crystal diffraction studies on LZIO, we expect an equally good crystal quality for it, given that all three crystals were grown using the same methods.

\section{Results }

RIXS measurements were performed to understand the low-energy elementary
excitations in all three compounds. RIXS at the Ir $\mathit{L_{\mathrm{3}}}$
edge, as a second order process involving x-ray absorption and emission
connected by an intermediate Ir-5$\mathit{d}$ state,\cite{Ament_11,KimYJ_18}
can sensitively probe the $\mathit{d}$-$\mathit{d}$ crystal-field
excitations,\cite{Gretarsson_13,Aczel_19} magnetic
excitations,\cite{KimJH_214,KimJH_327,Lu_17,Chun_18}
and even phonon excitations\cite{Meyers_18,Vale_19} in various iridates.

\begin{figure}
\centering{}\includegraphics[width=1\columnwidth]{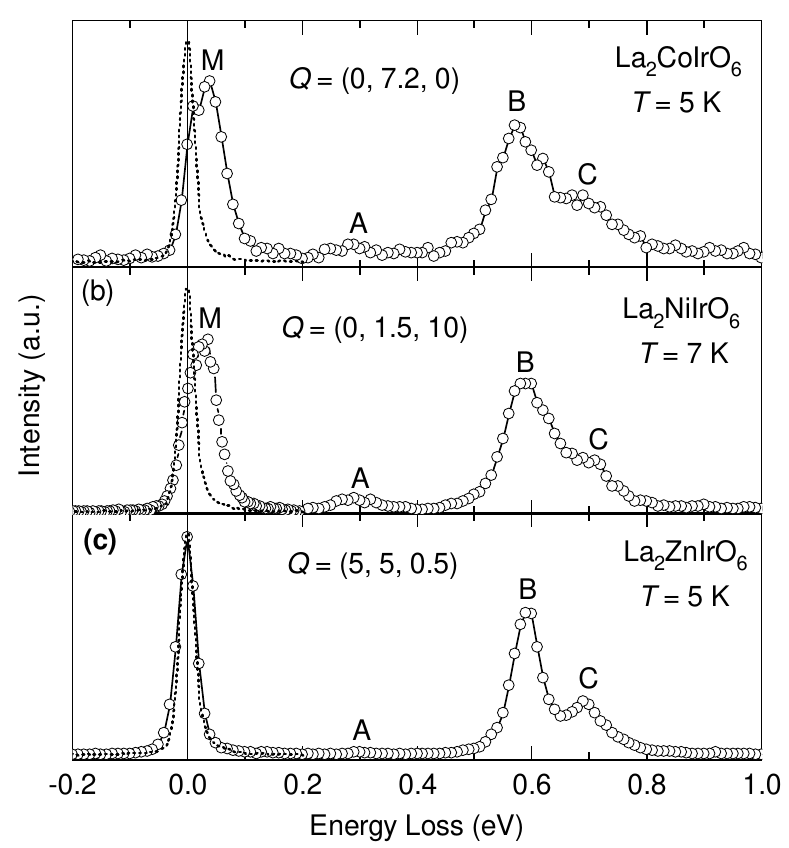}

\caption{The RIXS spectra with the energy transfer below 1 eV obtained at the
base temperature for LCIO (a), LNIO (b) and LZIO (c), taken at $\mathit{Q}$
= (0, 7.2, 0), (0, 1.5, 10) and (5, 5, 0.5), respectively. The dashed
lines represent the instrumental resolution.}
\end{figure}

Figure 1 shows the RIXS spectra in a wide energy transfer range below
1 eV obtained at the base temperature at $\mathit{Q}$ = (0, 7.2,
0), (0, 1.5, 10) and (5, 5, 0.5), for LCIO (a), LNIO (b) and LZIO
(c), respectively, revealing multiple common excitation features for
all three compounds, labeled A, B, and C. It is worth noting the similarity between all three spectra, indicating similar Ir electronic states. This observation is consistent with the recent hard x-ray photoelectron spectroscopy (HAXPES) study by Takegami et al., in which little difference between the Ir electronic structure for these three samples was observed.\cite{Takegami_20}
Our results are also consistent with the RIXS spectrum reported for LZIO in Ref. \onlinecite{Aczel_19},
where features B and C at the energy transfers of $\hbar\omega\sim$ 0.6$\mathds{-}$0.7
eV were attributed to the intraband $\mathit{t_{\mathrm{2g}}}$ crystal-field
excitations from the $\mathit{j_{\mathrm{eff}}}$ = 3/2 quartet to the $\mathit{j_{\mathrm{eff}}}$
= 1/2 doublet, seperated due to the noncubic crystal-field splitting
of the $\mathit{j_{\mathrm{eff}}}$ = 3/2 manifold at the Ir$^{4+}$ sites.
The origin of feature A around 0.3 eV is still not clear yet, which
might be related to the 3$\mathit{d}$-5$\mathit{d}$ magnetic hybridization,
as this mode is evidently stronger in LCIO and LNIO compared with
that in LZIO. A similar feature, attributed to 5$\mathit{d}$-4$\mathit{d}$ hopping, was observed in the recent RIXS study of Ag$_{3}$LiIr$_{2}$O$_{6}$.\cite{de la Torre_21}
We also note that a mode with a comparable energy ($\hbar\omega\sim$
0.42 eV) was observed in the RIXS spectra of Na$_{2}$IrO$_{3}$,
which was assigned to the particle-hole exciton across the charge
gap.\cite{Gretarsson_13} Note that these three materials are all semiconductors with similar gap sizes. The activation gap of 0.26 eV and 0.34 eV were estimated from the resitivity measurements on LCIO and LZIO, respectively.\cite{Narayanan_10,Gao_20} For LNIO, the density functional calculation estimates that the gap is also of a similar size around $\sim$ 0.2~eV.\cite{Wang_12} Therefore, peak A could also be due to the enhanced exciton contribution near the charge gap. These excitations above 0.2
eV are beyond the scope of this paper, and they will not be further discussed here.

Comparing the base-temperature RIXS spectrum of the three compounds
below 0.2 eV, it is noticeable that a low-energy mode around 40 meV
(labeled M) exists in both LCIO and LNIO, which is however absent
in LZIO. For LZIO, there is only a sharp elastic peak below 0.2 eV,
which is limited by the instrumental resolution (the dashed line).
To obtain some insights on the origin of the low-energy mode M, the
incident energy dependence of this mode was measured at 5 K at $\mathit{Q}$
= (0, 7.2, 0) for LCIO. As shown in Fig. 2, it exhibits a resonant
behavior around $\mathit{E_{i}}$ = 11.214 keV. As this energy is about 3~eV below the main Ir $\mathit{L_{\mathrm{3}}}$ absorption edge as
determined by the fluorescence line, we can conclude that the RIXS
process for M involves an intermediate state in which an Ir 2$\mathit{p_{\mathrm{3/2}}}$
core electron is excited into the Ir 5$\mathit{d}$ $\mathit{t_{\mathrm{\mathrm{2}g}}}$
states, instead of the 2$\mathit{p_{\mathrm{3/2}}}$ $\mathds{\rightarrow}$
5$\mathit{d}$ $\mathit{e_{\mathrm{g}}}$ transition. Since the lowest-energy
electronic excitations between the $\mathit{j_{\mathrm{eff}}}$ states occurs
with a much higher energy scale ($\mathds{\sim}$ 0.6 eV), the possible
origin of M around 40 meV can be either magnetic or phonon excitations,
both of which are expected to appear around this energy scale.

\begin{figure}
\centering{}\includegraphics[width=1\columnwidth]{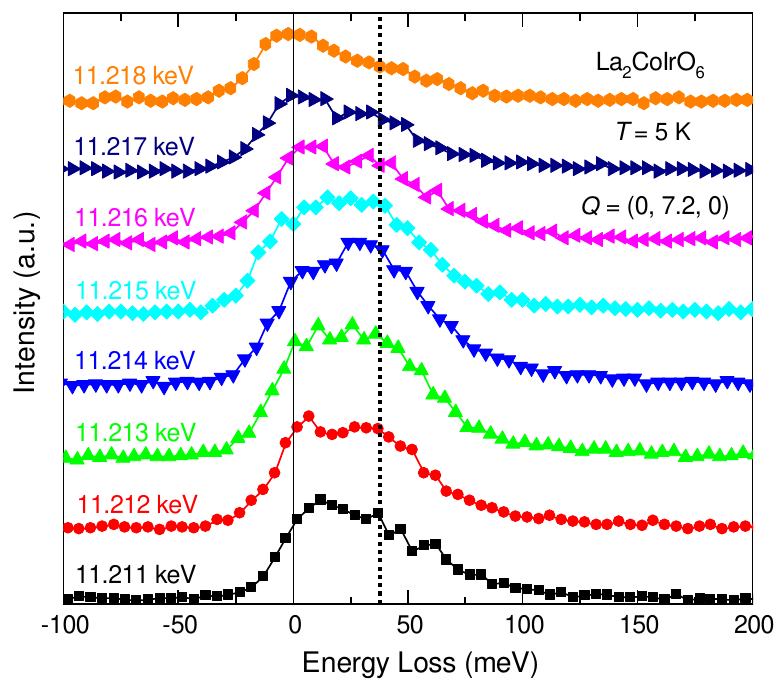}

\caption{The incident energy dependence of the RIXS spectrum of LCIO, taken
at $\mathit{T}$ = 5 K and $\mathit{Q}$ = (0, 7.2, 0). The dashed
line marks the low-energy mode M, with the maximal intensity achieved
at $\mathit{E_{i}}$ = 11.214 keV. }
\end{figure}

\begin{figure*}
\centering{}\includegraphics[width=1\textwidth]{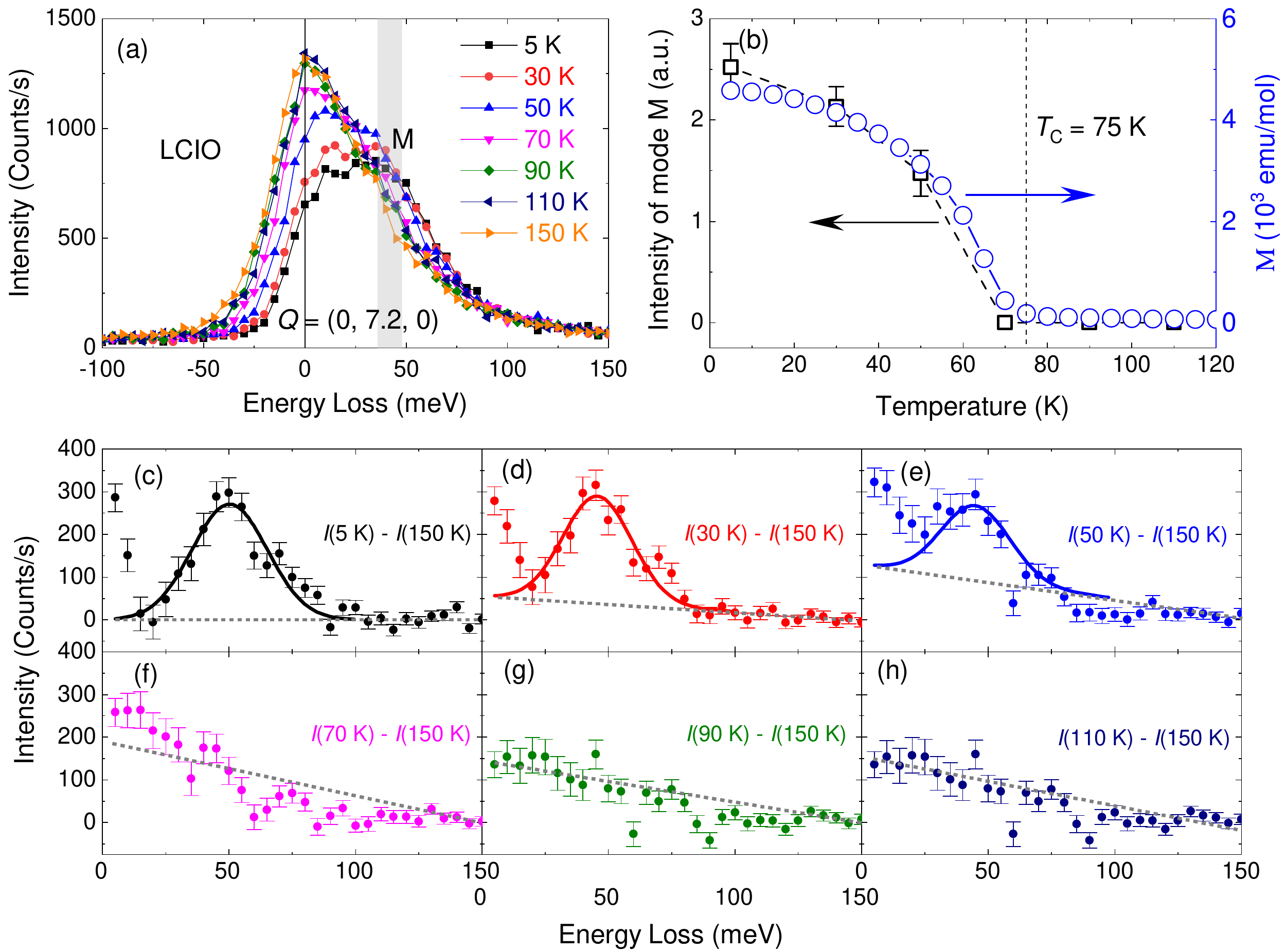}

\caption{(a) The RIXS spectrum of LCIO collected at different temperature at
$\mathit{Q}$  = (0, 7.2, 0), in which the shaded bar marks the
mode M. (b) The integrated intensity of mode M as a function of temperature,
as well as the dc magnetization of the LCIO single crystal
measured with an applied field of 0.2 T along the $\mathit{b}$ axis in the field-cooling process. (c)-(h) The
RIXS spectra at the Stokes side at 5 K (c), 30 K (d), 50 K (e), 70
K (f), 90 K (g) and 110 K (h), respectively, subtracted by that of
150 K after the Bose factor correction. The dotted and solid lines
present the sloping background and the fittings using a Gaussian profile.}
\end{figure*}

To further understand the origin of mode M, the RIXS spectrum of LCIO
at $\mathit{Q}$ = (0, 7.2, 0) is measured as a function of temperature.
As shown in Fig. 3(a), the amplitude of the elastic line increases
upon warming up, which is commonly observed in RIXS experiments and typically ascribed to thermal diffuse scattering.
In contrast, the intensity of mode M is suppressed
gradually with increasing temperature. The RIXS spectra for $\mathit{T}$ > 90 K are very similar, indicating
that the excitation spetrum does not change any more above $\mathds{\sim}$
90 K. Therefore, we decided to use the highest-temperature (150 K)
spectrum as a background to tease out the temperature dependence of
the mode M, after correcting the Bose population factor, $n(E)$=1/$(e^{E/k_{\mathrm{B}}T}-1)$, 
for the energy loss side. Note that the magnetic ordering temperature,
$\mathit{T_{\mathrm{C}}}$, is determined to be 75 K by dc magnetization measurements
(Fig. 3(b)). As shown in Fig. 3(c-h), below 70 K, the mode M with
the energy of $\mathds{\sim}$ 42(6)~meV becomes clearly visible after
such a correction and subtraction. With increasing temperature, this
mode fades away and gets buried into the sloping background finally,
strongly supporting its magnetic origin. This mode M can be fitted using a Gaussian profile superimposed on the
sloping background, by which its integrated intensity can be extracted.
A comparison between the intensity of mode M and the dc magnetization
of the LCIO single crystal measured in an applied field of 0.2 T (Fig.
3(b)) further corroborates our assignment of M as the magnetic
excitation. We also note that below 70 K, some
additional spectral weight remains below 20 meV even after the Bose
factor correction and background subtraction, whose origin is unclear
yet, although the incomplete subtraction is still a possibility. To clarify the nature of this additional feature, further RIXS
measurements with even higher resolution is needed.

The same procedure of data treatment for LNIO is also carried out.
As shown in Fig. S2 in the Supplemental Material,\cite{SM} the mode M in LNIO
displays a similar temperature dependence to that in LCIO, but it
persists only below 50 K. This is consistent with the fact that
the LNIO single crystal undergoes an antiferromagnetic-type transition
below $\mathit{T_{\mathrm{N}}}$ $\mathds{\sim}$ 42 K, as evidenced from the
magnetization data. After the Bose factor correction and background
subtraction, the energy of mode M in LNIO can be determined to be
$\mathds{\sim}$ 35(5) meV. Therefore, by combining the results from
both LCIO and LNIO, we can reach the conclusion that the mode M, appearing
in the RIXS spectra of both compounds, is clearly associated with
the magnetic excitations of ordered Ir$^{4+}$ moments.

\begin{figure*}
\centering{}\includegraphics[width=1\textwidth]{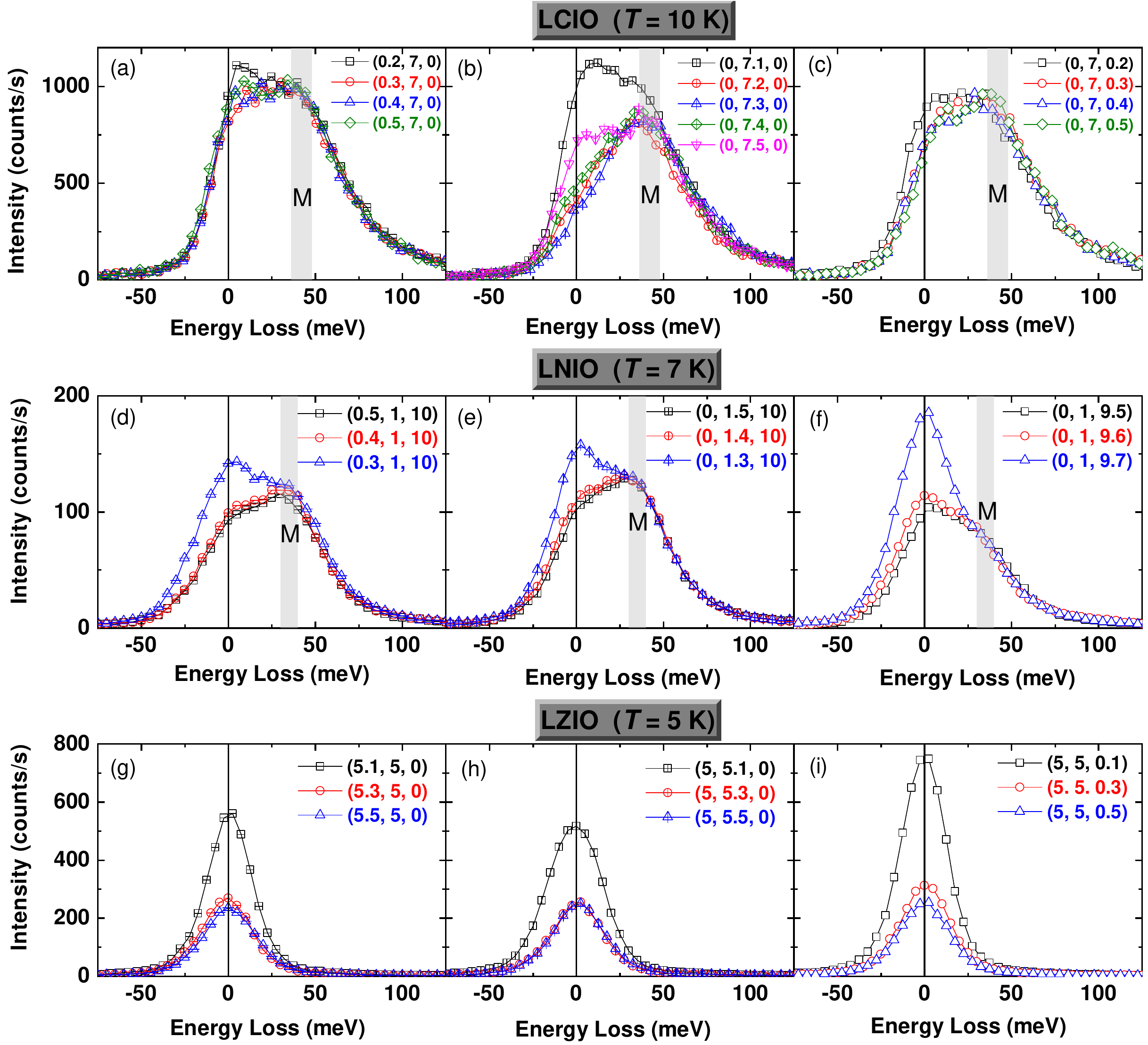}

\caption{The $\mathit{Q}$-dependence of the low-energy RIXS spetrum of LCIO
(a-c), LNIO (d-f), and LZIO (g-i), collected at the base temperature.
The shaded bars in (a-f) mark the magnetic excitation existing
in LCIO and LNIO. }
\end{figure*}

The $\mathit{Q}$-dependence of the low-energy RIXS spetrum of all
three samples along the reciprocal $\mathit{H,}$ $\mathit{K},$ $\mathit{L}$
directions were measured at the base temperature for LCIO (Fig. 4(a-c)),
LNIO (Fig. 4(d-f)), and LZIO (Fig. 4(g-i)). Note that a sizable temperature-independent
background is present in the RIXS spectra, as reflected by pretty
similar profiles for LCIO for $\mathit{T}$ > 90 K (see Fig. 3(a)).
Within the experimental uncertainty,
the energy of mode M stays almost constant, indicating an almost dispersionless
magnetic excitation for LCIO and LNIO with a very large excitation
gap of $\mathds{\sim}$ 42 meV and $\mathds{\sim}$ 35 meV, respectively.
In contrast, this feature is not observed for different
$\mathit{Q}$ values in LZIO with non-magnetic 3$\mathit{d}$ Zn$^{2+}$
ions, supporting that the observed mode M in LCIO and LNIO originate
from the 3$\mathit{d}$-5$\mathit{d}$ magnetic
hybridization. As the shaded bars in Fig. 4(a-f) show, the energy
of this magnetic excitation is higher in LCIO than that in LNIO,
probably indicating a larger magnetic energy scale in LCIO.

\section{Discussion And Conclusion}

We presented two compelling pieces of evidence for the magnetic origin of the observed non-dispersive excitation around 40 meV in LCIO
and LNIO. One is its temperature dependence, which tracks the magnetic order closely. The other is the fact that this mode is absent in LZIO, whose magnetic behavior is distinct from the other two compounds. Although tempting, it is not straigtforward to associate this with a magnon mode. The fact that the mode is almost dispersionless is unusual for a magnon mode. It is certainly possible that this excitation is composed of multiple magnon modes. With the energy resolution of the present RIXS instrumentation, we were not able to resolve any fine structure, which prevents us from comparing it with theoretical dispersion relations. Another possibility for this excitation could be a hybrid phonon-magnon mode. It is well known that optical phonon modes are fairly non-dispersive, and if phonon modes are strongly coupled to the magnon, the momentum and temperature dependence could be explained. In fact, the anomalous temperature dependence of the Raman $\mathit{A_{\mathrm{1g}}}$ mode in a DP Ba$_{2}$FeReO$_{6}$ was attributed to a strong coupling between the bond-stretching mode and magnetic order.\cite{Raman} Recently, Lee et al. also pointed out the existence of such a mode in LCIO based on their neutron powder  diffraction data.\cite{LeeS_18} These studies certainly support the idea of this mode arising from the magnetoelastic coupling. However, the difference in the excitation energy of about 15\% between LCIO and LNIO would be difficult to explain, since these two compounds have an almost identical structure and the phonon energies are expected to be similar.

The absence of mode M in our RIXS spectra of LZIO clearly indicates that
this magnetic excitation observed in LCIO and LNIO arises from
the hybridization between the unpaired 5$\mathit{d}$ and 3$\mathit{d}$
electrons, which was proposed previously to explain the x-ray magnetic
circular dichroism (XMCD) results of LCIO.\cite{Kolchinskaya_12}
The 5$\mathit{d}$--3$\mathit{d}$ hybridization was the subject of the recent x-ray absorption spectroscopy investigation\cite{LeeM_18} as well as the HAXPES study.\cite{Takegami_20} In particular, the latter study showed that the Ir-O hybridization is very strong in both $\sigma$ and $\pi$ symmetry. The DFT calculation in this study also shows strongly hybridized Ni and Co states near the Fermi level, which is absent for the Zn states in LZIO.\cite{Takegami_20} The strong antiferromagnetic interaction between Co/Ni and Ir in these compounds can be understood considering the oxygen mediated hybridization. Since Co and Ni can only have unoccupied states with minority spin direction, the parallel spin configuration for Ir would be forbidden (See Fig. 6 in Ref.~\onlinecite{LeeM_18}). Then one can understand the difference in the magnetic energy scale of LCIO and LNIO as arising from difference in hybridization due to slight structural difference. Specifically, the Ir-O-Co angle is $\sim 154^\circ$, while the Ir-O-Ni angle is $\sim 150^\circ$,\cite{Narayanan_10,Wang_12} meaning that the Ir-O-Co bond is closer to 180$^\circ$ that gives rise to stronger antiferromagnetic interaction.


For LZIO, in which
the 3$\mathit{d}$ shell of Zn$^{2+}$ is completely filled and there
is no medium of unpaired 3$\mathit{d}$ electrons, such a high-energy
magnetic excitation is absent. We note that magnetic excitations arising from Ir moments were observed around 2 meV in a recent inelastic neutron scattering study,\cite{Aczel_16} which is indistinguishable from the elastic line with the current RIXS resolution. It is interesting to note that despite different ground-state magnetic structure, LCIO and LNIO seem to have similar magnetic excitations due to antiferromagnetic superexchange-like coupling between Ir and Co/Ni.
Our results provides a solid evidence that
such a 3$\mathit{d}$-5$\mathit{d}$ magnetic coupling is playing
an imporant role in the magnetism of the DP iridates, similar to the
case in the DP rhenates.\cite{Plumb_13,Yuan_18} Further theoretical
studies on these materials will be of great importance to understand
the detailed magnetic interactions in them.

Another interesting observation is that from Fig. 3(b) and Fig. S2(b), the
feature M in either LCIO or LNIO disappears above
the magnetic ordering temperature. This is somewhat surprising, since classical spin systems described by a spin Hamiltonian typically show broadening and softening of the zone-boundary magnons, instead of a complete suppression as observed here.\cite{Huberman,Wan}
The observed suppression of magnetic mode above the transition temperature, however, is consistent with other recent studies on iridate materials. For example, a similarly dramatic temperature dependence was observed for a layered compound Sr$_3$Ir$_2$O$_7$ and a pyrochlore compound Eu$_2$Ir$_2$O$_7$.\cite{KimJH_327,Chun_18} These studies suggest that charge fluctuations are important in iridates and that the Hamiltonian beyond a simple spin model is required to account for the magnetic excitations in these materials.

In conclusion, the low-energy elementary excitations in a series of
double perovskite iridate single crystals including LCIO, LNIO and
LZIO are investigated by means of RIXS measurements at the Ir $\mathit{L_{\mathrm{3}}}$
edge. Clear magnetic excitations with a large excitation gap of $\mathds{\sim}$
42(6) meV and $\mathds{\sim}$ 35(5) meV were observed, for LCIO and
LNIO, respectively. This magnetic excitation disappears above
the magnetic ordering temperature, suggesting the inadequacy of a local spin model for describing magnetism of these double perovskite iridates.
The absence of such
a magnetic excitation in LZIO, in which the 3$\mathit{d}$ ions
are non-magnetic, suggests the importance of 3$\mathit{d}$-5$\mathit{d}$
hybridization in the magnetism of the DP iridates.

\bibliographystyle{apsrev} \bibliographystyle{apsrev}
\begin{acknowledgments}
The authors would like to acknowledge valuable discussions with Bo
Yuan and Shoushu Gong. Work at the University of Toronto was supported
by the Natural Sciences and Engineering Research Council of Canada,
through Discovery and CREATE program, and Canada Foundation for Innovation.
W.J acknowledges the support by the National Natural Science Foundation of China (Grant No. 12074023). C.J.W. was supported by the National Research
Foundation of Korea (Grant No. 2016K1A4A4A01922028 and 2020M3H4A2084417). Use of the Advanced Photon Source at Argonne National Laboratory is
supported by the U.S. Department of Energy, Office of Science, under
Contract No. DE-AC02-06CH11357.

\bibliographystyle{apsrev}
\bibliography{DP}
\end{acknowledgments}

\end{document}